# An ultracold low emittance electron source

G. Xia[a,b,*], M. Harvey[b,c], A. J. Murray[a,c], L. Bellan[d], W. Bertsche[a,b], R. B. Appleby[a,b], O. Mete[a,b], S. Chattopadhyay[a,b,c,e]

[a] *School of Physics and Astronomy, University of Manchester, Manchester, United Kingdom*
[b] *The Cockcroft Institute, Sci-Tech Daresbury, Daresbury, Warrington, United Kingdom*
[c] *Photon Science Institute, University of Manchester, Manchester, United Kingdom*
[d] *Department of Physics, University of Trieste, Trieste, Italy*
[e] *Department of Physics, University of Liverpool, Liverpool, United Kingdom*

E-mail: guoxing.xia@manchester.ac.uk

ABSTRACT: Ultracold atom-based electron sources have recently been proposed as an alternative to the conventional photo-injectors or thermionic electron guns widely used in modern particle accelerators. The advantages of ultracold atom-based electron sources lie in the fact that the electrons extracted from the plasma (created from near threshold photo-ionization of ultracold atoms) have a very low temperature, i.e. down to tens of Kelvin. Extraction of these electrons has the potential for producing very low emittance electron bunches. These features are crucial for the next generation of particle accelerators, including free electron lasers, plasma-based accelerators and future linear colliders. The source also has many potential direct applications, including ultrafast electron diffraction (UED) and electron microscopy, due to its intrinsically high coherence. In this paper, the basic mechanism of ultracold electron beam production is discussed and our new research facility for an ultracold, low emittance electron source is introduced. This source is based on a novel alternating current Magneto-Optical Trap (the AC-MOT). Detailed simulations for a proposed extraction system have shown that for a 1 pC bunch charge, a beam emittance of 0.35 mm mrad is obtainable, with a bunch length of 3 mm and energy spread 1 %.

# Contents

## 1. Introduction

Electron sources are the basic components for many applications ranging from electron diffraction [1] and electron microscopy systems [2], through fundamental electron scattering experiments [3], to state-of-the-art scientific research utilising particle accelerators [4]. These include x-ray free electron lasers, light sources and high-energy electron-positron colliders. For most contemporary applications and for modern particle accelerator facilities, electrons are typically created through either photoemission or through thermionic emission [5].

Ultracold atom-based electron sources have recently been proposed as an alternative to photo-emitters and thermionic-based electron sources [6-8]. The low electron temperatures associated with this kind of source (around 10 K or less) gives them an advantage over conventional electron sources that are typically several orders of magnitude hotter ($10^3$ K-$10^4$ K).

This lower temperature results in a beam with smaller volume and higher density in phase-space for a given bunch current. The beam emittance, which is defined by the volume the beam occupies in phase space, is hence improved by several orders of magnitude. In addition, cold electron sources have the potential to produce pulses with ultra-short bunch lengths (e.g. sub-picosecond), making them possible candidates for the next generation of particle accelerators. These include high luminosity colliders, plasma wakefield accelerators, injectors for free electron lasers and Compton backscattering techniques for production of x-rays. Further, since these new sources have much higher energy resolution, their coherence is correspondingly greater due to the low electron temperature. This makes them ideal for the development of a new generation of electron diffraction experiments, with potential use in structure studies ranging from materials science through to cell biology [9].

This paper reports on an ultracold electron source being developed using a novel alternating current Magneto-Optical Trap (AC-MOT). After this introduction, section 2 briefly discusses the mechanisms to produce ultracold atoms and plasmas. Some important quantities of the beam, such as brightness, emittance and transverse coherence, as well as their dependencies on the electron temperature, are given in section 3. Sections 4 and 5 present the working principles of the AC-MOT invented in Manchester, and discuss the ultracold electron source using this AC-MOT. Simulations for ultracold electron extraction and subsequent acceleration are then given in section 6.

## 2. Ultracold atoms and plasmas

Ultracold plasmas are produced through near-threshold photo-ionisation of laser-cooled and trapped atoms [10], since these atoms provide a high-density source of ultracold targets. The advent of reliable high stability continuous wave (CW) laser radiation has enabled rapid development of different methods to cool and trap atoms. Rubidium (Rb) is typically studied since these atoms have cooling and trapping transitions that are accessible with 780 nm laser radiation, a wavelength readily available from inexpensive diode lasers. Other targets can also be used, however they usually require more sophisticated and expensive lasers. The most common technique to cool and trap atoms employs a magneto-optical trap (MOT) [10,11], which consists of a red-detuned laser field in the presence of an inhomogeneous magnetic **B**-field (as provided by a pair of anti-Helmholtz coils). The MOT cools and spatially confines a cloud of $\sim 10^{11}$ atoms at micro-Kelvin temperatures. At these low temperatures the atoms are almost stationary, allowing virtually Doppler-free photo-ionisation to take place. The photo-ionising laser can be tuned near threshold to produce cold electrons and ions with very little excess kinetic energy. By carefully shaping the intensity profile of the laser beams, it is possible to produce cold electrons with a uniform distribution. This allows expansion of the electron cloud due to space charge to be countered by adopting standard electron lenses [9,12].

The excess kinetic energy produced in the photo-ionisation process (i.e. the laser energy above the ionisation limit) is mainly taken away by the electrons, due to the large mass difference between $Rb^+$ ions and electrons. Electrons with a well-defined temperature on the order of 10 K can be produced by finely tuning the ionisation laser frequency [13]. A simple model to understand the physical processes of formation of ultracold plasmas [10] considers a neutral charge distribution with the numbers of electrons and ions being equal, with a quasi-uniform spatial distribution when the atoms are initially photo-ionised. According to this model, since the electron cloud expands faster than the ion cloud, the central charge in the plasma becomes positive. The resulting ions create a Coulomb potential well, which traps electrons [14]. Electrons with energy greater than the well depth then escape, thereby increasing the depth. This process continues until no electrons remain that are capable of escaping the well, except those gaining energy through thermalisation and evaporation. The potential well finally becomes deep enough so that trapped electrons and ions co-exist independently, and an ultracold plasma is formed. The cold electrons can then be extracted by applying a strong electric field across this plasma.

Four-steps hence explain how cold electrons can be generated. (I) Atoms are laser cooled and trapped in a MOT (laser 1). (II) The cold atoms are excited to an intermediate state (laser 2). (III) A third laser (laser 3) then ionises the excited atoms only within the volume irradiated by lasers 2 and 3. In this way, the shape and density of the electron cloud is controlled. (IV) Cold electrons in the ultracold plasma are extracted by an externally applied electric field.

## 3. Emittance and brightness of the resulting electron beam

An ultracold electron source can be compared to photo-injector and thermionic electron sources, with its transverse brightness, $B_\perp$, which may be written as

$$B_\perp = \frac{I}{4\pi^2 \varepsilon_x \varepsilon_y} \quad (3.1)$$

where $I$ is the peak current of the beam, and $\varepsilon_x$ and $\varepsilon_y$ are the normalised transverse (horizontal $x$ and vertical $y$) beam emittances.

For conventional electron sources as used in particle accelerators, the best performing pulsed picosecond sources are RF photo-injectors, in which electrons are created by pulsed laser photoemission, followed by subsequent acceleration and bunch compression in RF fields [15]. In these sources, the normalised transverse emittance originates from various sources such as thermal effects, space charge, and RF field in the vicinity of the source. The dominating contribution comes from nonlinear space-charge forces [16] and this effect can be compensated downstream from the source by using a set of focusing elements. A recent study has also shown that the space-charge forces can be eliminated by a proper shaping of the radial intensity profile of a femto-second photo-excitation laser [17]. The RF contribution is induced inevitably due to the time varying nature of the field. The contribution from thermal effects depends on the size of the source (emission area), $\sigma_{source}$, and the temperature, $T$, as given in Eq 3.2 [13],

$$\varepsilon_{source} = \sigma_{source} \sqrt{\frac{kT}{mc^2}} \qquad (3.2)$$

where $k$ is Boltzmann constant, $m$ is the mass of electron and $c$ is the speed of light.

Thermal contribution to the total emittance can be reduced to less than an eV due to a low source temperature in a photoinjector. These contributions can be reduced further (by orders of magnitude) in a cold atom based electron source, as has been noted above.

The coherence length, $L_c$, is another important quantity for an electron source, in particular for electron diffraction applications. For an electron beam with a thermal distribution, the spatial coherence is given by [18],

$$L_c = \frac{\hbar}{\sqrt{mkT}} \qquad (3.3)$$

where $\hbar = h/2\pi$ is the reduced Planck constant. To observe interference in imaging applications, the coherence length must be at least twice the unit cell spacing of a sample. At the beam waist, the transverse spatial coherence length of the electron beam is given by [19],

$$L_c = \frac{\hbar}{mc} \frac{\sigma_r}{\varepsilon_r} \qquad (3.4)$$

where $\sigma_r^2 = \sigma_x^2 + \sigma_y^2$ is the transverse rms bunch radius and $\varepsilon_r^2 = \varepsilon_x^2 + \varepsilon_y^2$ is the transverse emittance. Eqs. (3.3) and (3.4) show that the transverse coherence is increased by reducing the electron beam temperature and therefore the beam emittance, as expressed in Eq. (3.2).

Current state-of-the-art electron sources increase brightness by reducing the emission area. As an example, photoelectron emission from needle cathodes have recently demonstrated high peak current of 2.9 A [20], with a source temperature typically of the order of $10^4$ K and an emission area of width 50 μm [21, 22]. Carbon nanotube (CNT) field emission sources reduce the emission area further [23], however these sources are limited to operation in the regime of very small (nA) currents due to their very small emission area, which is typically only a few nanometers in diameter.

The achievable peak current of a MOT based electron system is determined by many different factors such as loading rate, trapping efficiency and photoionisation efficiency. These are currently under study for the source at Manchester, so as to provide analytical and experimental characterisation. The peak current depends as much on the bunch length as it does on bunch population. In a photo-injector the initial bunch length is determined by the laser pulse width, which can have values as low as ~ fs with current technology. However in the cold atom based electron sources, the bunch length is a function of the laser width, atom cloud size and the details of the extraction system [24, 25]. As far as the lifetime of the electron source is concerned, there is almost no age limit for a MOT based cold electron source, since the atomic vapor injected into the trap can be continuously replenished. The quantum efficiency of these new sources is also comparable to metallic-based photocathode materials, such as copper or magnesium [26].

As an example for later comparison, the Accelerator Test Facility (ATF) photo-injector at Brookhaven National Lab can produce 0.5 nC electron bunches with $I \approx$ 120 A and $\varepsilon_x \approx$ 0.8 mm mrad [27]. This corresponds to a beam brightness of $B_\perp = 5 \times 10^{12}$ A/m$^2$rad$^2$. The corresponding value for the cold atom based electron source will be given in section 6.

## 4. The alternating current Magneto-Optical Trap (AC-MOT)

A new type of magneto-optical trap has been invented in Manchester, which uses *alternating current* to generate the magnetic trapping field, in combination with high-speed polarization switching of the damping laser field. This new atom trap, known as an AC-MOT, allows the magnetic fields within the trapping region to be switched off over 300 times faster than is possible in a conventional DC-MOT [28]. The AC-MOT has the significant advantage that the magnetic **B**-fields required for trapping are eliminated prior to the production and extraction of low energy photoelectrons. This has two significant advantages. The first is that the hyperfine sub-states of the target prior to laser excitation and ionisation are unperturbed by any **B**-field, and so are degenerate in energy. A larger fraction of cold atoms can then be excited to an intermediate state by laser 2, the energy resolution of this state being given by its lifetime and any power broadening due to the laser (we assume here that Doppler effects are eliminated due to the low temperature of the trapped atoms). Ionizing laser 3 then produces photoelectrons whose kinetic energy resolution is given by that of the intermediate state and that of the ionising laser. Klar and co-workers pioneered this type of work in 1992 using a metastable target [29], where they produced photoelectrons with energy resolution as low as 0.05 meV. In their work they detail the importance of eliminating external electric and magnetic fields to achieve this resolution. Modern laser systems have much lower bandwidths than those used in [29], and so it is expected that the new electron source being developed here will improve on this resolution.

The second advantage of eliminating the trapping **B**-field is that the resulting low energy photoelectrons can then be extracted and accelerated using simple electrostatic lenses. Accurate modeling programs can be used to describe the trajectory of the electrons, without the need to apply corrections due to the spatially varying trapping **B**-field. Understanding these initial electron trajectories is critical for accurate modeling of subsequent acceleration stages that are required to produce an electron beam.

It has been demonstrated that the AC-MOT trapping fields can be switched on and off at up to 10 kHz, so that the cold atoms remain within the trapping region during the photo-ionisation process. The AC-MOT uses a sinusoidally varying magnetic field in combination with trapping lasers whose polarization is switched from $\sigma^+$ to $\sigma^-$ at the same rate. This combination produces a cold atomic ensemble of equivalent density to that of a conventional DC-MOT [28]. The AC-MOT switches the trapping **B**-field off when it passes through zero, thereby eliminating transient (eddy) currents in surrounding conductors, and their associated fields. Laser photo-ionization and subsequent extraction of low energy electrons is then carried out when the **B**-field is zero. Following electron extraction, the AC-magnetic field is switched on again, and the cycle repeated. The AC-MOT remains fully functional even with a 50% duty cycle, so that electrons can be extracted at rates up to 5 kHz, allowing quasi-CW production of cold electrons from the source.

By contrast, significant eddy currents are induced when a conventional DC-MOT is rapidly switched to zero, the fields associated with the induced currents taking ~10-20 ms to decay to a level where low energy electrons can be extracted and manipulated. This delay results in substantial trap losses, leading to a low repetition rate for experiments. The advantages of the AC-MOT when using low energy electrons are clearly demonstrated in [28], which shows the first electron impact ionisation cross section measurements from cold atoms, for a range of incident electron energies from ~2 eV to ~70 eV. These experiments were not possible using a DC-MOT due to trap losses. Such effects were completely eliminated using the AC-MOT.

## 5. The cold electron source in Manchester

By virtue of the significant advantages in yield, resolution and repetition rate, a new cold electron source based on the AC-MOT is currently being built in Manchester. Atoms cooled and trapped in the AC-MOT will be selectively photo-ionised with a pulse from one of the laser systems in the Photon Science Institute (PSI). Cold electrons will then be extracted by applying an electrostatic field. After extraction, the electron beam will drift a short distance (up to several tens of centimeters), and will be measured using a microchannel plate (MCP). A pepper pot [30], together with YAG or OTR screen measurement system will be used to characterise the beam emittance. The beam emittance will then be used to derive the effective electron temperature and to study the dynamics of cold beam formation.

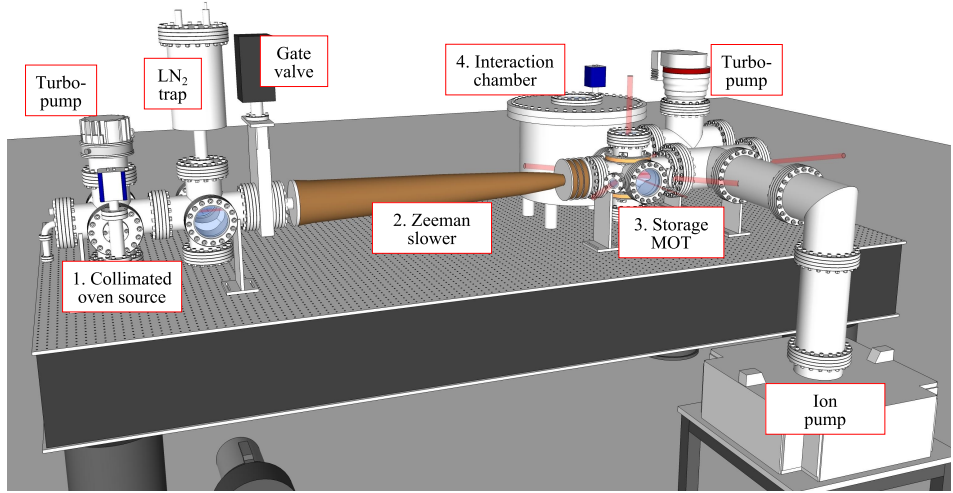

**Figure 1.** Schematic of the new cold electron source in Manchester, showing (1) the source chamber, (2) the Zeeman slower, (3) the storage MOT chamber and (4) the interaction chamber where the AC-MOT and electrostatic optics will be located.

Figure 1 shows the apparatus under construction. A rubidium oven together with a Zeeman slower provides a continuous high flux of atoms that load a storage DC-MOT [31]. Cold atoms from the storage MOT will be guided to the interaction chamber, where they will be re-trapped in the AC-MOT. Here they will be photo-ionised, and a cold electron beam will be extracted, accelerated, focused and characterised.

Rb will be produced as a collimated atomic beam from the oven, before injection into the Zeeman slower and storage MOT. The thermal velocity profile of the atomic beam is reduced and compressed inside the Zeeman slower, so that atoms injected into the storage MOT have a low velocity (~ 40 m/s). This is less than the capture velocity of the MOT, so that these slow Rb atoms rapidly accumulate in the trap (typical trap filling times using these techniques are much less than 1 second, with ~$10^{11}$ atoms accumulating in the trap in this time [28]). The Rb atoms rapidly cool in the storage MOT to close to the photon recoil limit (~ 140 μK for Rb). Cold atoms from the storage MOT will then be injected into the AC-MOT using a low power pushing laser beam, when required. The efficiency of this injection process depends upon the background vacuum, distance between storage MOT and AC-MOT, and characteristics of the pushing laser beam. The background pressure in the system has already been demonstrated to be less than $10^{-10}$ torr, and so losses due to background gas collisions are expected to be negligible during transport. By carefully designing the pushing laser beam to minimise inter-atom collisions, the injection efficiency into the AC-MOT is then expected to be high (up to 85% efficiency has been demonstrated for this process using a laser beam red-detuned from resonance by 1.2 GHz [32]).

The use of a storage MOT located separately from the AC-MOT has two advantages. The first is that the DC magnetic fields from both the Zeeman slower and storage MOT can be reduced to very low levels in the AC-MOT, by using magnetic shielding. The second is that cold atoms can be accumulated continuously in the storage MOT from the Zeeman slower, for selective injection into the AC-MOT during the retrapping cycle (i.e. when the AC-MOT **B**- field is on). In this way, the AC-MOT is continuously fed with cold atoms, allowing the yield of extracted cold electrons to be maximised. The steady-state yield of photoelectrons will then depend upon the loading rate of the AC-MOT and the rate of ionization of cold atoms.

The saturation density and structure of the cold atom cloud in the AC-MOT depends on the shape of the magnetic field, on detuning from resonance of the trapping laser beams, and on radiation trapping within the cold atom ensemble. Typical densities in the AC-MOT are found to be $\sim 10^{11}$ atoms/cm$^3$ [28], and it has been demonstrated that clouds of cold atoms with shapes ranging from spherical densities through to thin sheets of atoms can be produced [33].

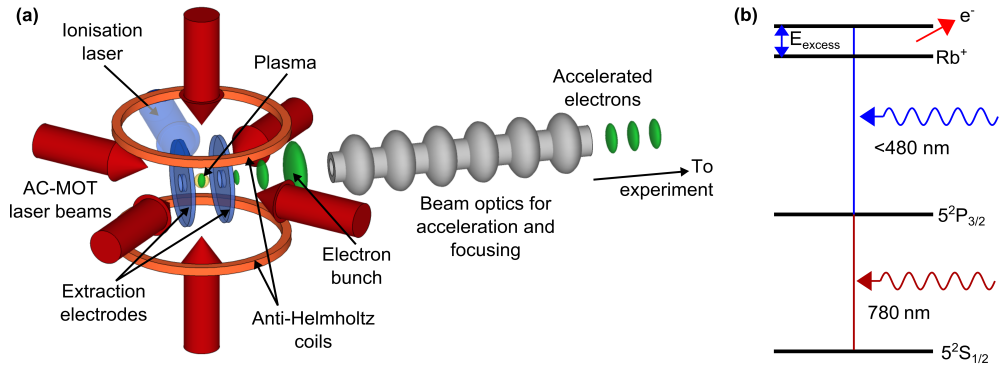

**Figure 2.** (a) Schematic of the cold electron beam extraction and transportation (not to scale), (b) The stepwise photo-ionization process.

Once atoms are cooled and trapped in the AC-MOT as shown in Figure 2(a), the **B**-field will be switched off and the excitation and ionising laser beams will be injected. The Rb ionisation potential is $\sim 4.18$ eV, so direct photo-ionisation from the ground state requires a laser with a wavelength less than ~296 nm. As an alternative, radiation at less than $\sim 480$ nm can be used to photo-ionise Rb atoms that are resonantly excited to the 5P state using radiation at $\sim 780$ nm, as in Figure 2(b). Both processes will be explored in these experiments. The ionizing beams at 296 nm and 480 nm will be sourced from frequency doubled CW and pulsed lasers, whereas the 780 nm radiation will be produced from a CW Ti:Sapphire laser. A suite of high resolution tunable CW and pulsed lasers are available in Manchester, and these will be used to determine the optimum conditions for photo-ionisation of trapped atoms. The ionisation efficiency depends upon both the intensity and wavelength of the laser beam (estimates of this efficiency are currently being determined using the existing source in Manchester [28]). For direct ionisation from the ground state at $\sim 296$ nm, it is this efficiency that determines the photoelectron yield. For the stepwise process, where atoms are excited to the 5P state followed by ionisation at $\sim 480$ nm, at most 50 % of the cold atom ensemble can be in the excited 5P state, and so the photoelectron yield will correspondingly reduce. It is however easier and less expensive to produce radiation at 480 nm compared to 296 nm, and so the stepwise process may prove to be the optimal (and most cost effective) method of electron production.

An electrostatic field will be used to extract and accelerate cold electrons from the AC-MOT, as shown in Figure 2(a). After the beam is extracted, it will be transported to the acceleration stage for further shaping and characterization. In principle, a simple two-electrode or three-electrode system can be used to extract cold electrons produced in the AC-MOT. The simulations results for the extracted beam presented in section 6 consider a three-electrode system with given apertures. By adopting electrodes constructed from high transparency tungsten mesh (95% transmission), optical access for the laser beams can be assured. Unlike the cold atom trap in [28] which housed the anti-Helmholtz coils inside the vacuum chamber, the AC-MOT in this experiment uses trapping coils located outside the chamber. Consequently, the electrostatic field across the trapping region is only governed by the configuration of the electrodes. As a first test, an extraction system similar to that of a Pierce grid will be used to separate the photo-ionisation and acceleration stages. This is advantageous due to smaller electric field heating effects on the cold electrons created at the center of the AC-MOT. The main acceleration section following extraction will then have a high electric field for fast acceleration of the emerging electrons. In this region the cold electrons can be further compressed by an RF cavity so as to avoid beam emittance expansion, after which they will be transported and focused into a target/imaging system.

The extraction and acceleration system currently being built will be used as a test bed to design a low emittance, cold electron source for future applications. An ultracold electron diffraction facility is planned as an initial application for this source. The source also has potential use in high-energy electron diffraction experiments. It further has prospects of providing a low emittance, high brightness electron source for injection into free electron lasers, linear colliders and plasma wakefield accelerators etc. In this case, a significant increase in intensity (~2 orders of magnitude) would be required. These may be realized through development of different technologies to increase the electron yield of the system, e.g. by increasing the loading rates and achieving high atom numbers, or by tuning the lasers and downstream extraction and accelerating fields to optimize the beam distribution.

## 6. Design of the beam extraction and acceleration electrodes

Following formation of the ultracold plasma in the AC-MOT, an electrostatic field will extract and accelerate the cold electrons to tens of keV, as shown in Figure 2. As noted above, the charge of the electron bunch relies on the density of cold atoms in the AC-MOT, which depends upon the flux of atoms from the source and Zeeman slower. The electron pulse length depends on the spatial overlap between excitation and ionization laser beams, as well as on the time/space profile of the applied electric potential of the accelerator. For a detailed understanding of the cold electron beam specifications, several methods for generating a DC electric field for extraction and acceleration of electrons have been modeled.

As a first effort to extract the beam efficiently, a three-electrode system was designed. Electrostatic fields were modeled in a cylindrically symmetric fashion using the Poisson/Superfish code [34]. One sample case is presented in Figure 3. The cold atoms and subsequent plasma are centered between the first two electrodes (biased at − 32 kV and − 11 kV, respectively) so that the electric field is relatively low (5.850 kV/cm in the interaction point). The third accelerator plate (the horn, biased at 0 kV) forms a stronger second stage of acceleration (8.905 kV/cm, which has a weak focusing effect on the electron bunch due to the radial component of the electric field). This design potentially minimises emittance growth of the electron beam and reduces the accelerating field that is required, thereby simplifying the fast switching requirements of the accelerator.

The potentials from the three electrodes are illustrated in Figure 3. The resulting fields are well below the breakdown limit of the vacuum, which we conservatively assume to be 30 kV/cm [35]. The distance between the first two electrodes is about 4 cm. The third electrode (horn) is designed to extract the beam with high efficiency and good beam quality. The electric field between the middle electrode and the horn electrode is about 25 kV/cm.

The field map in Figure 3 is calculated by assuming cylindrical symmetry within a rectangular volume. One side corresponds to the *z*-axis and has Von Neumann boundary conditions, while the remaining three edges adhere to Dirichlet boundary conditions. The electrodes are assumed to be perfect conductors. The system is open in the transverse directions, to allow optical access for the laser beams.

The beam parameters, such as beam size, bunch length, beam emittance and energy spread are calculated using the GPT (General Particle Tracer) code [36]. This is a simulation platform for the study of charged particle dynamics in electro-magnetic fields. In the GPT code a set of macro-particles are used to simulate electrons and ions. The calculated electric field from Poisson/Superfish is used as input for the external electric field. All routines concerning Coulomb interactions between these particles are derived from the electric field interactions between the $i^{th}$-particle and all other particles as well as the external electric field. The interactions between particles correctly model the space-charge effects of the bunch.

A cylindrically symmetric bunch of $10^4$ macro-particles has been used in the simulations. The overall density distribution is Gaussian-like in the transverse planes ($x, y$) with a bunch radius of 2 mm, and it peaks along the $z$-axis at $z = 0$ with an initial bunch length of 0.6 mm, as shown on the upper left plot in Figure 4 at $z = 0$ mm. This kind of distribution can be experimentally obtained, and shows the reasonable behavior of the bunch specifications.

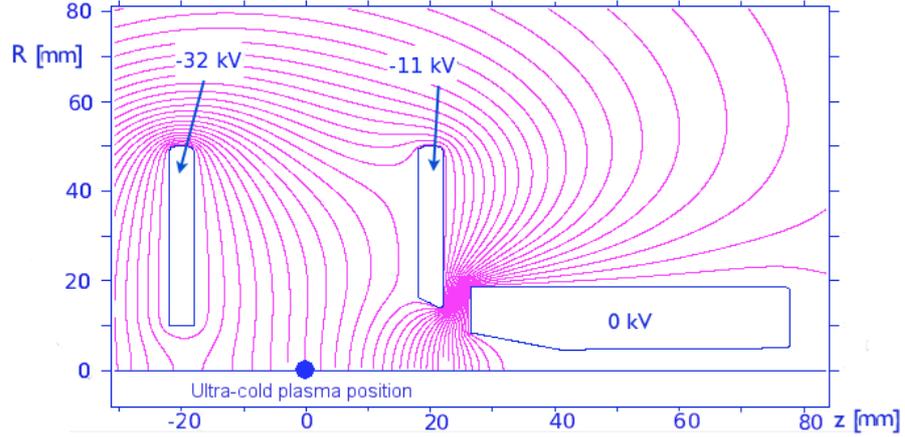

**Figure 3**. Contours of electric potential plotted for voltages applied to the extraction electrodes as shown. The field map is generated using Poisson/Superfish. Cylindrical coordinates are used and they are symmetric with respect to the horizontal axis. The horizontal axis is the $z$-axis (i.e., the beam transport line axis) and the vertical axis is $R$, which corresponds to the radial coordinate. The curved lines represent equipotential lines.

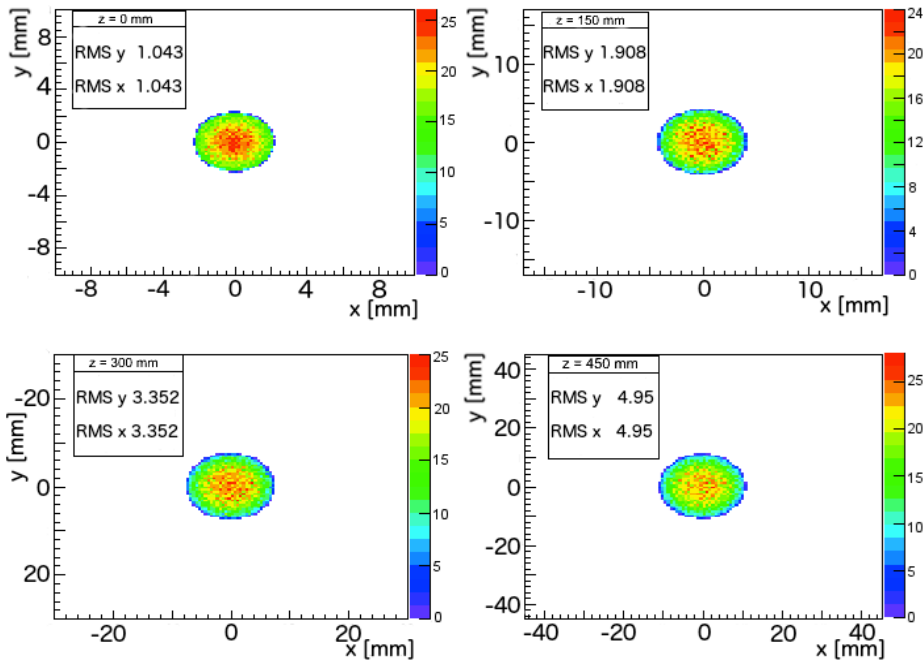

**Figure 4.** Snapshots of the bunch transverse density distributions ($x, y$) at different locations of the beam line along $z$-axis. The beam area is rescaled in order to compare their distributions. A canvas with the RMS of the bunch (in mm) is also shown in each plot.

Figures 4-9 show GPT simulation results for a 1 pC bunch after exiting the center of the cold atom cloud and being transported 500 mm in the longitudinal $z$ direction. This bunch-charge is considered to be an upper limit for the source currently being built, and so provides a test of the effects of space charge on the resulting beam. We assume that after 500 mm beam transport, it will be possible to employ additional magnetic lenses such as quadrupoles or solenoids to focus and control the beam size and emittance.

Figure 4 gives the bunch transverse ($x, y$) density distribution at different locations, i.e., at $z = 0$ mm, $z = 150$ mm, $z = 300$ mm and $z = 450$ mm, respectively. The beam area is rescaled in the four plots in order to compare their distributions. The RMS values are also shown in each plot. This figure shows that after the beam is extracted and transported downstream from the AC-MOT, the particles in the bunch expand and redistribute themselves due to space-charge effects. The plot in the lower right shows that the resulting particle distribution is almost uniform at about z = 450 mm.

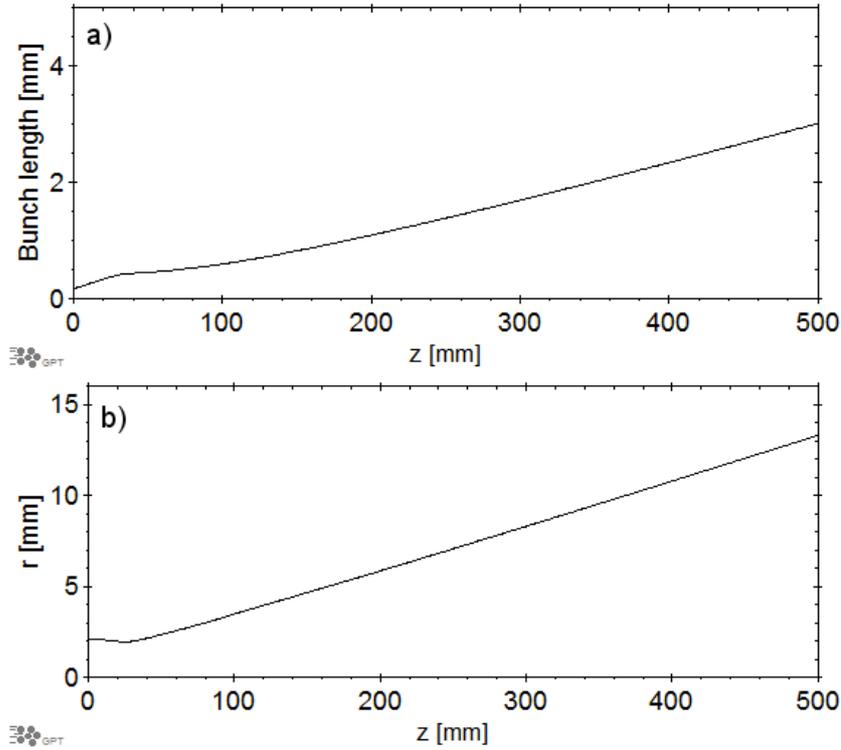

**Figure 5.** a) Maximum beam radius for a bunch of 1 pC charge against the beam transport axis $z$. b) Bunch length of a 1 pC charge against the beam transport axis $z$.

The bunch length evolution is shown in Figure 5-a), where it is seen that the bunch length increases linearly to about 3 mm after 500 mm transport. Figure 5-b) gives the maximum beam radial size against the beam transport direction $z$. It indicates that the beam size increases almost linearly along the beam transport direction after exiting the centre of the cold atom cloud.

Figure 6 compares the beam emittance evolution for 1 pC and 10 fC charges per bunch, with respect to the beam transport direction $z$. This figure indicates that the beam emittance increases significantly with the bunch charge when the beam energy is low (i.e. immediately after extraction). This is mainly due to space-charge. After two-stages of acceleration the emittance shows two growths. A constant behavior is found when the beam is inside the horn, which is 50 mm long. After exiting the horn, the emittance increases by a small amount and then a constant behavior is found due to a bunch redistribution process. This is followed by a slow monotonic decrease of the emittance, until it reaches an asymptotic value of 0.35 mm mrad. To achieve a lower emittance, the bunch charge can be reduced, or the electric field can be increased between the extraction electrode plates. It is seen that the emittance is markedly reduced, becoming 0.19 mm mrad at 500 mm for the 10 fC case since the effects of space charge are reduced in this case. By increasing the electric field to close to the vacuum breakdown limit (~ 30 kV/cm), we further predict a beam emittance of 0.07 mm mrad. This is comparable to data obtained from other groups [35, 37, 38].

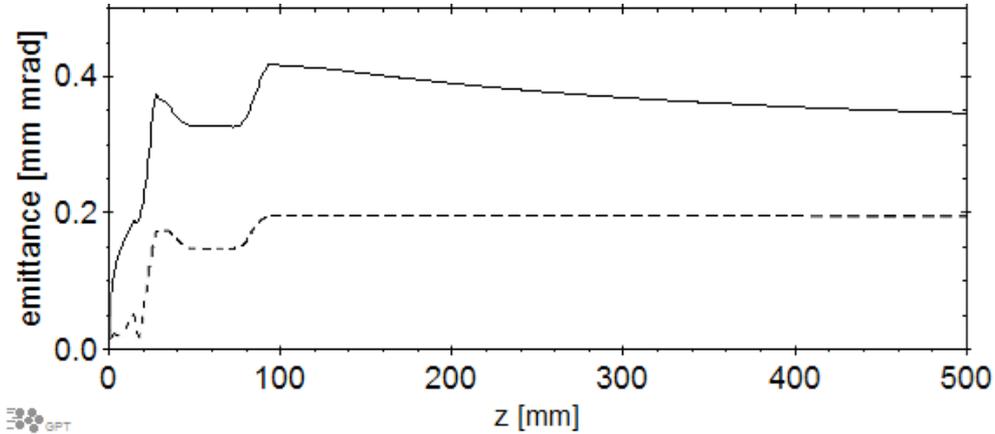

**Figure 6**. Transverse beam emittance for a bunch with charges of 1 pC (solid line) and 10 fC (dashed line) as a function of beam propagation axis.

After 500 mm, the beam emittance is 0.35 mm mrad, and so for a bunch length of 3 mm and extracted charge of 1 pC the brightness is estimated to be of the order of ~ $10^{10}$ A/m$^2$ rad$^2$. The value of bunch length chosen here depends primarily on the atom cloud radius. Whereas a 1 pC bunch charge is used in these initial simulation studies, this depends on the loading rate and ionisation efficiency. It is worth noting that it is possible to get higher brightness electron beams by carefully controlling different variables, such as the flux and density of the atom cloud and the shape of the trapping and ionization lasers and extraction fields [13].

The AC-MOT being used here has a further significant advantage, since it provides closer control over the electron extraction parameters due to elimination of the magnetic fields that are present in a DC-MOT. An example of this advantage is given in figure 7, which shows the calculated emittance, by using GPT, from both an AC-MOT and DC-MOT operating under the same trapping conditions. The magnetic field from the DC-MOT is found to significantly affect the predicted emittance of the electron beam, compared to that from the AC-MOT where electrons are only extracted when the magnetic field is zero. These calculations clearly show that use of the AC-MOT will allow production of electron beams with increased brightness and coherence, due to this significantly lower emittance.

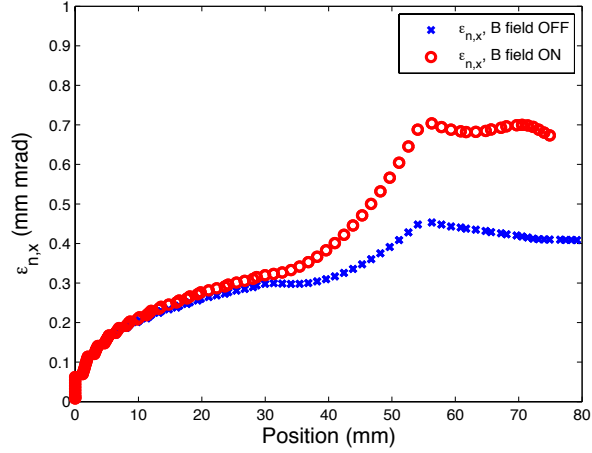

**Figure 7.** Calculated effect of the magnetic field in the trapping region on the normalised transverse beam emittance. The advantage of the AC-MOT (with a magnetic field of zero) compared to that of the DC-MOT is clearly seen.

From the model for the AC-MOT source, the transverse coherence length (Eq. 3.4) is found to be $L_c$ = 1.46 nm, which is sufficient for imaging structures on a nanometer scale [24]. This reveals that the average beam energy is 21.90 keV with a standard deviation of 0.24 keV, corresponding to an energy spread of 1.1 %. This spread is thought to originate from the initial momentum spread of the atom cloud, and also from differences in uniformity of the electric field in the extraction region. The extraction system is under further optimisation to minimise this spread in energy.

Additional magnetic focusing can be applied after the extraction region to control the beam size. Initial simulations using a solenoidal field are shown in figure 8, which depicts the maximum beam radius for a 1 pC bunch along the beam transport axis $z$. In this simulation, we assume the magntic field from the solenoid is 100 Gauss along the z-direction, and that the solenoid of length 100 mm is positioned 325 mm downstream from the center of the AC-MOT. The figure shows that the beam radius can be significantly reduced using this field, so that the minimum beam radius is ~ 0.07 mm at $z$ = 470 mm. It is therefore advantageous in the new facility to implement a solenoidal field to increase the density of the electron beam.

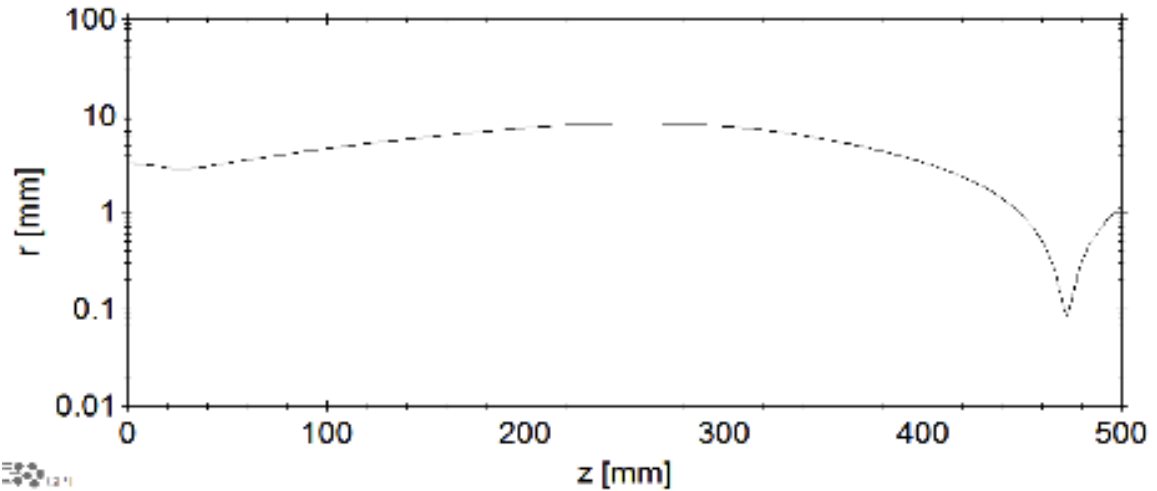

**Figure 8.** Maximum beam radius for a bunch of 1 pC charge against the beam transport axis $z$, using a 150 mm long solenoid centred 325 mm from the extraction region which produces a magnetic field of 100 Gauss within the solenoid.

It is clearly important to establish if the solenoidal focusing field modeled in figure 8 affects the emittance of the beam produced from the extraction region. Figure 9 demonstrates that under the conditions described above, the emittance in the extraction region is not affected in any significant way by this field. However in the case of a high field, or for a solenoid located closer to the extraction region the emittance will increase. Under such conditions compensation coils can be located around the extraction region to cancel any residual fields.

Alongside these preliminary simulations, a more susbstantial focusing system is under study to control the beam envelope and to reduce the emittance due to space-charge [39]. This scheme implements two solenoids, one to provide focusing and a second to cancel the magnetic field at the extraction region. The results from these simulations will be presented in a future publication.

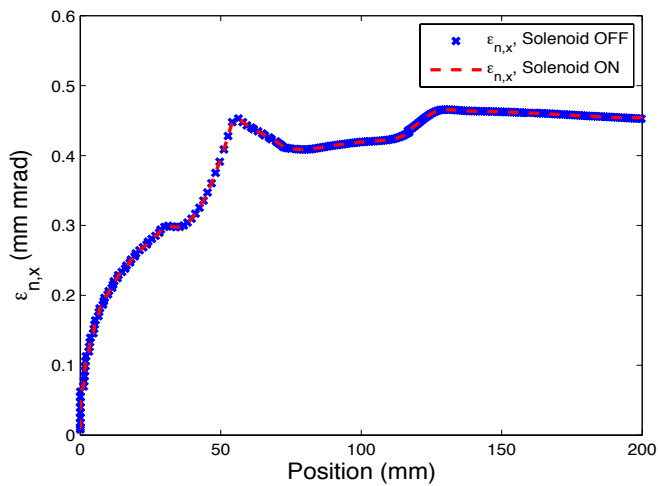

**Figure 9.** Emittance in the extraction region when the downstream focusing solenoid is on and off.

## 7. Conclusions

Ultracold electron sources have the advantage of low beam emittance and high brightness, compared to conventional electron sources used in current particle accelerators [40, 41]. The simulations presented here reveal that an electron source based on the AC-MOT in Manchester should achieve a 1 pC charge bunch, with a beam emittance of 0.35 mm mrad and bunch length of 3 mm for an average energy of ~ 22 keV at the exit of the extraction region After 500 mm beam transport, the resulting beam brightness in this case is about ~$10^{10}$ A/m$^2$ rad$^2$. The predicted high coherence length ($L_c$ = 1.46 nm) should then allow imaging experiments to be performed on nanometer scale structures. To increase the brightness RF fields may be adopted, potentially enabling a higher bunch charge and lower emittance. The ultracold electron source project in Manchester will study the details of this cold electron beam production, extraction and characterisation. These new ultracold electron sources are currently a worldwide initiative, and are expected to revolutionise electron source technology. They will then find many applications in the future.

## Acknowledgments

The authors would like to thank E. J. D. Vredenbregt, S. B. van der Geer for illuminating discussion on this project. This work is supported by the STFC, Cockcroft Institute and the University of Manchester strategy fund AA11718.